# Real-Time Recognition of In-Place Body Actions and Head Gestures using Only a Head-Mounted Display

Jingbo Zhao[1*]    Mingjun Shao[1]    Yaojun Wang[1†]    Ruolin Xu[2]

College of Information and Electrical Engineering, China Agricultural University[1]

Department of Electrical and Computer Engineering, Duke University[2]

**ABSTRACT**

Body actions and head gestures are natural interfaces for interaction in virtual environments. Existing methods for in-place body action recognition often require hardware more than a head-mounted display (HMD), making body action interfaces difficult to be introduced to ordinary virtual reality (VR) users as they usually only possess an HMD. In addition, there lacks a unified solution to recognize in-place body actions and head gestures. This potentially hinders the exploration of the use of in-place body actions and head gestures for novel interaction experiences in virtual environments. We present a unified two-stream 1-D convolutional neural network (CNN) for recognition of body actions when a user performs walking-in-place (WIP) and for recognition of head gestures when a user stands still wearing only an HMD. Compared to previous approaches, our method does not require specialized hardware and/or additional tracking devices other than an HMD and can recognize a significantly larger number of body actions and head gestures than other existing methods. In total, ten in-place body actions and eight head gestures can be recognized with the proposed method, which makes this method a readily available body action interface (head gestures included) for interaction with virtual environments. We demonstrate one utility of the interface through a virtual locomotion task. Results show that the present body action interface is reliable in detecting body actions for the VR locomotion task but is physically demanding compared to a touch controller interface. The present body action interface is promising for new VR experiences and applications, especially for VR fitness applications where workouts are intended.

**Index Terms**: Human-centered computing—Human computer interaction (HCI)—Interaction paradigms—Virtual reality; Human centered computing—Human computer interaction (HCI)—HCI design and evaluation methods—User studies

## 1 INTRODUCTION

Body actions and head gestures have been introduced to VR systems as natural interfaces for interaction with virtual environments. One application of body action recognition is for WIP techniques [1], [2], in which the locomotion velocity, direction and actions are estimated from the tracking data captured from users using trackers and are mapped to virtual environments to generate virtual walking experiences. Body actions associated with WIP techniques typically include stepping in place, jogging in place, jumping and strafing, etc. The introduction of avatars for user body representation also requires in-place body action recognition algorithms for WIP techniques when full-body tracking is not available for ordinary VR users to render avatar animation. For example, when a user is stepping in place, jogging in place or jumping, etc., correct animation clips have to be played on the avatar to reflect a user's body movements based on the recognition results.

Existing approaches to recognizing in-place body actions designed for WIP techniques often require specialized hardware and/or additional trackers, which typically include Kinect sensors for skeleton tracking [3], smart insoles that can detect foot pressure [4] and wearable sensors to track specific body joints [5], etc. However, many ordinary VR users only possess an HMD. The use of specialized hardware for body action recognition makes it less likely for VR designers and engineers to consider using body actions as interfaces for interaction experience in VR. Consequently, VR end-users would not have the opportunity to access body action interfaces.

Head gesture recognition has been previously explored by several studies. It has been demonstrated that head gestures can be used for answering Yes/No questions in virtual environments [6] and for selection and browsing in augmented reality (AR) [7]. However, existing work only considered recognizing head gestures using the tracking data from an HMD. No previous study considered a unified approach to recognizing body actions and head gestures using only head-tracking data from an HMD. This also potentially hinders the exploration of the use of body actions and head gestures for new interaction experiences in virtual environments and with avatars. On the other hand, if we intend to recognize in-place body actions using only head tracking data of an HMD, it is necessary to incorporate a mechanism to recognize head gestures. Otherwise, head movements, such as head tilting, are likely to be wrongly recognized as certain body actions (e.g., head tilting left may be wrongly recognized as body strafing left, etc.).

To tackle these problems, we proposed a unified 1-D CNN network that can recognize ten types of in-place body actions during walking and eight types of head gestures while standing still (see Figure 1 and Figure 2 below) using only head tracking data from an HMD. We considered this network a body action interface (head gestures included) and evaluated its utility and user preference through a virtual locomotion task, in which participants were asked to start their task by nodding their head, and avoid obstacles by jumping, squatting and strafing during in-place walking. To compare this body action interface, we ran the same locomotion experiment with a conventional Oculus touch controller interface as the control condition. Participants were asked to complete the same locomotion task using the buttons and the thumbstick on the controller interface to control an avatar's movements for comparison.

The goal of the present study is to propose an algorithm that can recognize body actions and head gestures using only head tracking data from an HMD and evaluate its utility and user preference through a user study. The primary contributions of the present study are two-fold:

(1) We present a unified two-stream 1-D CNN network capable

J. Zhao and M. Shao contributed equally to this work.
*e-mail: zhao.jingbo@cau.edu.cn
†e-mail: wangyaojun@cau.edu.cn

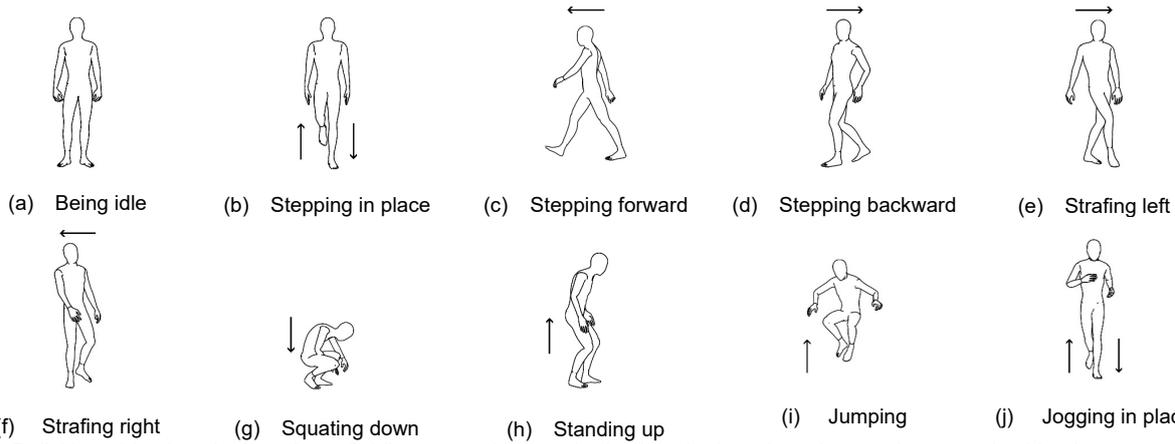

Figure 1: Definition of in-place body actions. Subfigures (a)-(j) show the in-place body actions that can be recognized by the two-stream 1-D CNN network.

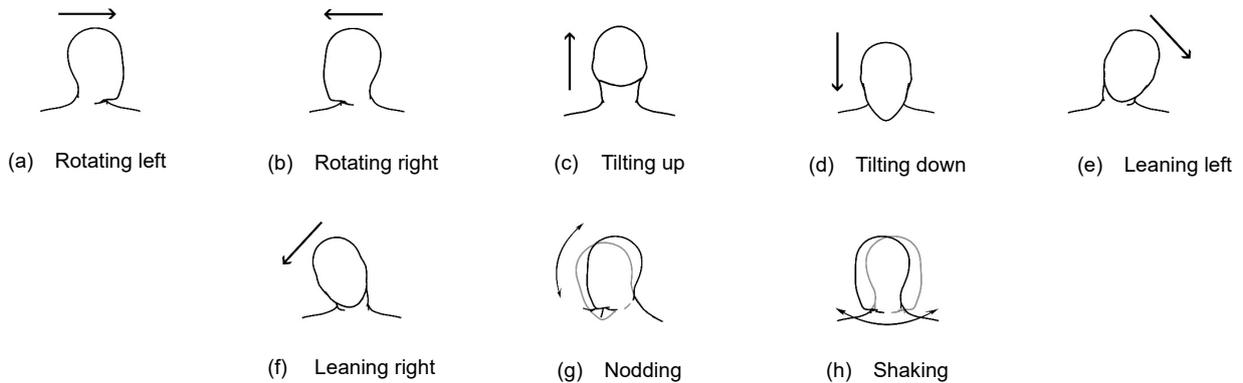

Figure 2: Definition of head gestures. Subfigures (a)-(h) show the head gestures that can be recognized by the two-stream 1-D CNN network.

of recognizing ten types of in-place body actions and eight types of head gestures.

(2) We present a virtual locomotion task that evaluated this 1-D CNN network as a body action interface in real-time and compare it to a conventional interface based on the Oculus touch controller.

Apart from using this body action interface for interactive experiences in virtual environments, it can also be used for VR locomotion behavioral analysis. Interfaces that can recognize the type and the duration of body actions and head gestures performed by participants can be helpful tools for VR researchers. Without such interfaces, experimental data have to be manually analyzed.

## 2 RELATED WORK

In this section, we review previous work on in-place body action recognition and head gesture recognition.

### 2.1 In-Place Body Action Recognition

Previous research has shown that it is possible to detect gait events from head motion data tracked by an HMD. Using the linear velocity data tracked from an HMD, a 1-D CNN network was designed to detect whether a user was standing or walking during in-place walking and the goal was to lower the stop latency of WIP techniques [8] through this stop detection algorithm. It has also been shown that apart from standing and walking, more gait phases including mid stance, double limb unsupported, and opposite mid stance can be recognized using a 1-D CNN network from the yaw data and head positions given by an HMD [9]. The detected gait phases can be used to render pre-defined leg animations that are associated with specific gait phases to reflect a user's leg movements during in-place walking [9]. However, more complex body actions or gestures have to be recognized using specialized hardware and/or additional trackers. For example, smart insoles were installed in shoes to detect foot pressure and a Long Short-Term Memory (LSTM) network was designed to recognize foot gestures [4]. Another approach was to attach HTC trackers on a user's knees to provide additional tracking data in addition to the data provided by an HMD. These data were considered point clouds and a point-cloud CNN network was designed to recognize in-place gestures [5]. Researchers have also investigated the classification of locomotion speed gains based on foot tracking data given by HTC trackers using traditional machine learning algorithms [10]. In addition, Kinect sensors have been widely used for various interaction tasks [11]. It has also been used for the recognition of body actions for WIP techniques [3]. We comment that using the tracking data from an HMD itself is sufficient for recognizing a number of in-place body actions and head gestures, other specialized hardware and/or additional trackers may not be necessary as we demonstrate in our study.

### 2.2 Head Gesture Recognition

An earlier research has been conducted to recognize head gestures based on the tracking data from an inertial measurement unit (IMU) placed on a person's head [12]. A thresholding technique was adopted to recognize head gestures based on the geometric features and the statistical features extracted from head tracking data. However, the application of the proposed algorithm was not demonstrated in this work. Since VR and AR based HMD devices also have the capability to track head motions using a hybrid optical-inertial approach. Researchers have proposed algorithms to recognize head gestures based on the tracking data from HMD devices and demonstrated their utility in VR and AR contexts.

Typical algorithms to recognize head gestures based on HMD devices include dynamic time warping (DTW) with k-nearest neighbor classifiers (KNNs) [13], DTW features and geometric features with support vector machines (SVMs) [7] and cascaded hidden Markov models (HMMs) [14]. The applications of these algorithms include user authentication in AR [13], menu selection and general browsing activities in AR [7], and answering Yes/No questions in VR [6]. To date, no previous work considered including in-place body action recognition algorithms to a head gesture recognition framework to enable in-place body action recognition.

Vision based approaches to recognizing head gestures have also been proposed by researchers. In classical vision based approaches, head pose needs to be estimated from a captured video stream first [15]–[18]. The estimated head pose changes over time can be considered as time series data for classification using state machines [16] or using machine learning algorithms, such as SVMs [17] and HMMs [15], [18]. More recent work has explored the use of deep learning algorithms to recognize head gestures, in which head gestures are directly inferred from video streams that capture a person's head [19], [20]. The application of vision based head gesture recognition has been demonstrated for computer dialog box responses [17]. Vision based approaches are also important for human-computer interaction and video analysis, but in the context of VR and AR, in which devices are generally inside-out systems [21] nowadays, outside-in systems [21] using external cameras to capture a user's head for head gesture recognition are less preferred due to the requirement of camera setup and calibration.

## 3 METHODS

### 3.1 Hardware and Software of the VR system

The Oculus Rift S headset is used in our study for data collection and the user study. It is a tethered HMD powered by a high-performance computer to process tracking data and generate high-quality graphics. It features a six degree-of-freedom (DOF) inside-out tracking system, which consists of five cameras mounted on the HMD. Tracking data are fused with inertial measurement data to estimate the HMD's position and orientation in the physical environment [22]. This HMD is accompanied by two 6-DOF Oculus touch controllers, which are equipped with infrared diodes, tracked by cameras on the headset, and inertial sensors fitted inside the controllers. The tracking data given by the HMD and its touch controllers include position $(x, y, z)$, linear velocity $(\dot{x}, \dot{y}, \dot{z})$, linear acceleration $(\ddot{x}, \ddot{y}, \ddot{z})$, Euler angles $(\psi, \theta, \Phi)$, angular velocity $(\dot{\psi}, \dot{\theta}, \dot{\Phi})$ and angular acceleration $(\ddot{\psi}, \ddot{\theta}, \ddot{\Phi})$, in a sampling frequency of 80 Hz. These data can be accessed through the C/C++ API from the Oculus PC SDK or through blueprints from the Unreal Engine. We only make use of the tracking data from the headset without using the tracking data from the touch controllers or other hardware to design and implement our two-stream 1-D CNN network to recognize body actions and head gestures. The host computer for this study is equipped with an Intel i5-10500 CPU, 32 GB memory, and an Nvidia Geforce 1660s graphics card with 6 GB graphics memory, running a Windows 10 64-bit operating system. Unreal 4.26 was used to design and test the proposed approach.

### 3.2 Definition of In-Place Body Actions and Head Gestures

We defined ten types of in-place body actions (shown in Figure 1) for in-place walking and eight types of head gestures (shown in Figure 2) for static stance. In-place body actions include: (1) being idle (standing still); (2) stepping in place; (3) stepping forward; (4) stepping backward; (5) strafing left; (6) strafing right; (7) squatting down; (8) standing up; (9) jumping; (10) jogging in place. Head gestures include: (1) rotating left; (2) rotating right; (3) tilting up; (4) tilting down; (5) leaning left; (6) leaning right; (7) nodding; (8) shaking; These are most of the in-place actions and head gestures that a user can perform during in-place walking. We did not include rotating right and rotating left for in-place body actions as these can be confused with head rotating left and head rotating right. This remains a challenge to be solved in a future study.

Table 1: Statistics of short sequences of body actions and head gestures for training and testing.

| Class Label | Total | Training | Testing |
|---|---|---|---|
| Being Idle | 32807 | 24704 | 8103 |
| Stepping in place | 16549 | 12260 | 4289 |
| Stepping forward | 7175 | 5224 | 1951 |
| Stepping backward | 9201 | 6846 | 2355 |
| Strafing left | 5414 | 3962 | 1452 |
| Strafing right | 6378 | 4794 | 1584 |
| Squatting down | 8144 | 6157 | 1987 |
| Standing up | 7784 | 5818 | 1966 |
| Jumping | 8099 | 6053 | 2046 |
| Jogging in place | 25732 | 19229 | 6503 |
| Rotating left | 7848 | 6150 | 1698 |
| Rotating right | 8001 | 6147 | 1854 |
| Tilting up | 7584 | 5849 | 1735 |
| Tilting down | 7454 | 5823 | 1631 |
| Leaning left | 7000 | 5275 | 1725 |
| Leaning right | 6954 | 5433 | 1521 |
| Nodding | 29599 | 22040 | 7559 |
| Shaking | 30357 | 22715 | 7642 |

### 3.3 Dataset Collection and Preparation

We developed a data collection application using Unreal 4.26 with the Oculus blueprints and C/C++ programming to collect the head tracking data from the Oculus S headset. A commercial Unreal plugin called VR Spectator Control Window (available from the Unreal Marketplace) was used to help create a control panel with buttons. In total, thirty students (age: 18-28, 15 males and 15 females) from China Agricultural University were invited as volunteers to participate in the data collection. We collected eighteen types of action data, defined in Section 3.2, for each participant. For each type of action, we performed four collection trials and each trial collected 4-s tracking data, during which the participant performed the action that was asked. The procedure followed the one described in the previous work by Zhao and Allison [14]. During a collection session, a participant wore the Oculus Rift S headset and stood in front of the host computer. For each type of body action, a researcher pressed the corresponding button on the control panel (visible only to the researcher on the computer monitor). Immediately, a prompt indicating the type of the action that a participant needed to perform was shown (visible to both participants in the view of the headset and the researcher on the computer monitor). A countdown timer was also started at the same time to count from four to zero by seconds. A participant was expected to complete the required action within 4 s with their preferred movement speed. Labeling of the action was done at the same time by the collection application within the 4-s interval. For actions other than standing still, we further processed each 4-s data sequence through thresholding to remove the parts of the sequence that corresponded to standing still when a user had not initiated an action or had already completed the required action. We then converted all data sequences into short sequences of length 40 with a stride of 1. These corresponded to running sequences of 0.5 s given a sampling frequency of 80 Hz, which are buffered using queues during the real-time operation of the proposed two-stream 1-D CNN network. For dataset partition, short sequences that belong to the first three collection trials of all participants were used as the training set while the short sequences of the fourth collection trial were used as the testing set. In Table 1, we present the total numbers of the short sequences that belong to the training set and the testing set with body action types and head gesture types.

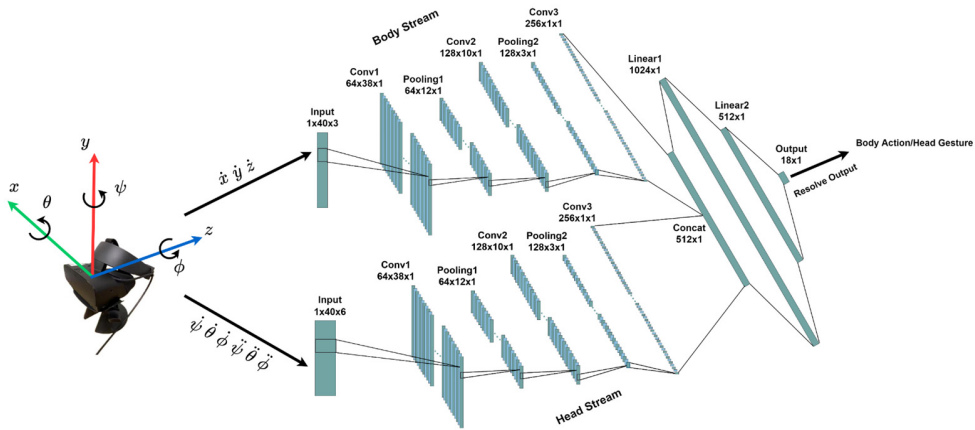

Figure 3: Two-stream 1-D CNN network for in-place body action and head gesture recognition. The upper stream is the body stream for body action feature extraction. It takes tracked linear velocity from the Oculus Rift S to extract body action features. The lower stream is the head stream for head gesture feature extraction, which takes angular velocity and angular acceleration to extract head gesture features. The extracted features are concatenated after the third convolutional layer of both streams, followed by two fully connected layers and an output layer for classification. The output is an 18-D vector. The final output is resolved by selecting the class label with the highest predicted value.

### 3.4 Unified Two-Stream 1-D CNN Network for Body Action and Head Gesture Recognition

In Figure 3, we present the architecture of the unified two-stream 1-D CNN network for body action and head gesture recognition. The upper body stream is used to extract body action features while the lower head stream is for the extraction of head gesture features. The body stream takes only linear velocity $(\dot{x}, \dot{y}, \dot{z})$ from the Oculus Rift S headset. Recall that the collected tracking data have been converted to short sequences of length 40, which correspond to a buffered queue of 40 elements during real-time operation, the dimension of the input data to the network is thus 1×40×3. Three convolutional layers with channels of 64, 128 and 256 are used for feature extraction. The selections of filter size, stride and padding for all three convolutional layers are 7, 1 and 2. The activation function is ReLU. Max pooling layers with a kernel size of 3 are applied after each of the first two convolutional layers. The structure of the head stream is almost the same as the body stream, except that it takes angular velocity $(\dot{\psi}, \dot{\theta}, \dot{\Phi})$ and angular acceleration $(\ddot{\psi}, \ddot{\theta}, \ddot{\Phi})$ to extract head gesture features. The dimension of input data to the head stream is 1×40×6. The settings of filter size, stride and padding are the same as the body stream. The ReLU activation function is used and max pooling layers with a kernel size of 3 are also applied as in the body stream. The output from the last convolutional layers of the body stream and the head stream are concatenated. The dropout is set to 0.2. Classification is performed using two fully connected layers with sizes of 1024 and 512. The output layer gives 18-dimensional predicted values as the output of the network. The final predicted body action or head gesture is resolved by selecting a body action or a head gesture with the highest predicted value. The loss function was selected to be cross-entropy. The optimizer was Adam. The proposed two-stream 1-D CNN network was implemented using PyTorch 1.8.2 CPU version with Python 3.8. Training of the network was performed with a batch size of 512, a learning rate of 0.0001 and a training epoch of 60, which yielded an accuracy of 96.16%. The confusion matrix that shows the classification performance is given in Figure 4. For the real-time application of the network, PyTorch was interfaced with Unreal 4.26 through the C/C++ interface, using an Unreal plugin called SimplePyTorch available from Github. The trained PyTorch model was loaded into memory by C/C++ before the VR application based on Unreal 4.26 started to process head-tracking data. The GPU version of PyTorch was not supported to be interfaced with Unreal 4.26 and so it was not considered.

We next discuss our motivation to design this network. Recognition of body actions and head gestures is often considered a problem of classifying and recognizing time series data. Approaches to tackling this problem included using HMMs [23] and LSTMs [24]. Recently, 1-D CNNs have been shown to be another promising approach for time series data classification and recognition [25], [26]. Given our setting of the body action and head gesture recognition problem and three different approaches, it was unclear which one is the best without evaluating their performance using collected HMD tracking data. In addition, in-place body actions are more related to linear motions and head gestures are more related to angular motions. Since we wished to decouple body action features and head gesture features, we considered linear velocity and linear acceleration for body action recognition and considered angular velocity and angular acceleration for head gesture recognition. We then developed three different algorithms based on 1-D CNNs, HMMs and LSTMs to evaluate their performance on body action recognition using linear velocity, linear acceleration and both. Head gesture recognition was evaluated using angular velocity, angular acceleration and both. The 1-D CNNs have three convolutional layers with sizes of 64, 128 and 256. The first two convolutional layers with ReLU as the activation function, are followed by max pooling layers with a size of 3 (as in the head stream and the body stream in Figure 3). Following the third convolutional layer are the two fully connected layers with sizes of 1024 and 512 and an output layer with a size of the number of body actions or the number of head gestures. The LSTMs are designed with three hidden layers, each with a hidden size of 256, followed by two fully connected layers with sizes of 1024 and 512. An output layer with a size of the number of body actions or head gestures follows the fully connected layers. LSTMs and 1-D CNNs were implemented using PyTorch 1.8.2. LSTMs were trained with 25 epochs while the 1-D CNNs were trained with 60 epochs, all with a learning rate of 0.0001. The HMM models used the K-Means for vector quantization. The number of symbols and the number of states of HMM models were swept to find the best combination of these two parameters that gave the highest accuracy, with a training epoch of 500. HMMs were implemented using the hmmlearn 0.2.7 package. The results of classification performance are given in Table 2, which shows that 1-D CNNs outperformed LSTMs and HMMs in each combination of data streams. Thus, 1-D CNNs were selected as our choice to implement our two-stream architecture. In addition, using linear velocity with an accuracy of 96.37% has the close performance to a combination of linear velocity and linear acceleration with an accuracy of 96.69% for body action recognition with 1-D CNNs. For head gesture recognition, a combination of angular velocity and angular acceleration with an accuracy of 95.50% performed better than using these two data streams alone (AngVel: 94.83%, AngAcc: 88.62%) with 1-D CNNs. Thus, we decided to use only linear velocity for body action recognition and a combination of angular velocity and angular acceleration for head gesture recognition. CNNs apply the same filter coefficients to data sequences whereas LSTMs and HMMs determine the present state based on their previous state when evaluating data sequences. CNNs are generally more efficient in computation than these two algorithms. Considering that our data sequences are fixed length and there is no need to learn long term dependency from data sequences, CNNs are better suited to the present task than LSTMs and HMMs.

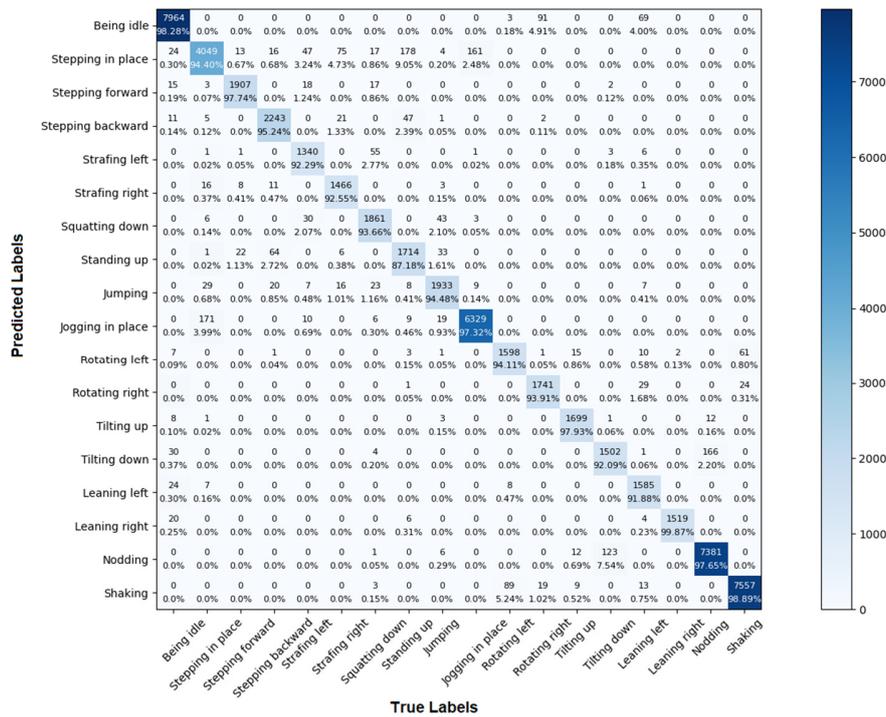

Figure 4: Confusion matrix of the classification results on the testing set using the proposed two-stream 1-D CNN network. The confusion matrix was normalized over the true (columns) condition. The horizontal axis represents the true labels, and the vertical axis represents the predicted labels. Each cell element represents the proportion of the number of the predicted labels to the total number of the true labels, with the number of short sequences included in each cell. The diagonal elements represent correctly classified outcomes. All other off-diagonal elements along a column are wrong predictions.

The second problem was how to fuse the output from two 1-D CNNs, which respectively recognize body actions and head gestures, to give a single output. Here, we tested two approaches. The first approach was to use an SVM [27] to fuse the predicated values from two 1-D CNN networks. The second was to use three fully connected layers to fuse the output. Using the same short-sequence data for training and testing, we found that the second approach, which used three fully connected layers for classification with an accuracy of 96.16%, outperformed the SVM approach, which gave an accuracy of 95.10%. Thus, fusion using three fully connected layers was finally adopted and the architecture of the two-stream 1-D CNN network was finalized, as shown in Figure 3. To confirm that the proposed two-stream network is more accurate than a single stream 1-D CNN network that uses linear velocity, angular velocity and angular acceleration for body action recognition and head gesture recognition, we implemented this single stream 1-D CNN and trained and tested the network. An accuracy of 95.68% given by this single stream network confirms the effectiveness of the proposed two-stream network, which has an accuracy of 96.16%. Compared to the single stream network, the two-stream network has two dedicated branches to extract translational features and rotational features, respectively, which enables it to have better performance in terms of accuracy.

Table 2: Classification performance of 1-D CNNs, LSTMs and HMMs on body actions and head gestures, with different combinations of data streams. AngVel denotes angular velocity, LinearVel denotes linear velocity, AngAcc denotes angular acceleration and LinearAcc denotes linear acceleration. Best values for each algorithm are in bold.

|  | Data Streams | 1D-CNNs | LSTMs | HMMs |
| --- | --- | --- | --- | --- |
| **Body Action** | LinearVel | 96.37% | 95.46% | **91.51%** |
|  | LinearAcc | 94.14% | 91.81% | 72.08% |
|  | LinearVel + LinearAcc | **96.69%** | **95.59%** | 77.93% |
| **Head Gesture** | AngVel | 94.83% | 94.06% | **89.60%** |
|  | AngAcc | 88.62% | 84.79% | 76.07% |
|  | AngVel + AngAcc | **95.50%** | **95.30%** | 75.02% |

As shown by the confusion matrix in Figure 4, the accuracy of each class cannot reach 100%. To ensure reliable recognition of body gestures and head gestures, we monitor the recognized body actions and head gestures during the real-time operation of the network. Whenever we find that five consecutive recognized body actions or head gestures are the same, we determine that a body action or a head gesture is reliably recognized. The final output is then given by the network and is used for interaction with virtual environments. Otherwise, the network returns a label -1 that represents an invalid body action or an invalid head gesture. This is considered a five-sample monitoring procedure.

There are three additional design considerations included to make the classification more reliable for real-time operation. First, when people intend to jump, they normally bend their legs a bit before jumping. This is recognized by the network as squatting first, followed by jumping. This is an unwanted recognition result as it triggers the avatar to slide on the ground during moving when a user bends their legs preparing to jump. But the user's actual intention is to jump over obstacles. Thus, we monitor the vertical positions of the HMD of the buffered 40 elements. Whenever we recognize squatting or jumping from the network, we calculate the first difference of the vertical positions of the buffered 40 elements. Then, for jumping, we sum those differences that are positive. When the sum is larger than 10 cm, we confirm that the user intends to jump. Similarly, for squatting, we sum the negative differences. When the sum is smaller than -10 cm, we confirm that the user intends to squat. Second, shaking is likely to be confused with head rotating left and rotating right, and nodding can be confused with tilting up and tilting down during preliminary real-time testing. This can also be observed from the last two columns of the confusion matrix on nodding and shaking for offline training and testing. Thus, we apply thresholds to shaking $\tau_s$ ($\tau_s = 120$) and nodding $\tau_n$ ($\tau_n = 75$) such that only when the predicted values of shaking and nodding pass their thresholds, we decide that shaking or nodding are recognized. This also means that only motions with large magnitudes are recognized. The thresholds were determined using the training samples with the trained two-stream 1-D CNN network, with manual adjustment. For example, training samples of rotating left, rotating right and shaking were given to the trained network as inputs to obtain their predicted values from the network. The threshold was determined to be the maximum difference between the predicted values of rotating left and right and shaking. Manual adjustments were performed to obtain desired recognition result. This technique was also applied to find the threshold for tilting up, tilting down and shaking. Third, a threshold was also applied to jogging in place ($\tau_j = 75$) to differentiate it from stepping in place. The threshold was also determined using the process described above for nodding and shaking. These three considerations were incorporated into the output resolution stage before the five-sample monitoring procedure to help determine the final recognized body action or head gesture. The end-to-end latency values for recognizing body actions and head gestures were estimated using the testing dataset. For body actions that do not have thresholds, the mean and the standard deviation of the estimated latency are 0.56 s $\pm$ 0.07 s. For jogging with the threshold $\tau_j$, the estimated latency is 1.15 s $\pm$ 0.55 s. For head gestures except nodding and shaking, the estimated latency is 0.57 s $\pm$ 0.07 s. For nodding and shaking, with thresholds of $\tau_n$ and $\tau_s$, the latency values are 1.03 s $\pm$ 0.80 s and 1.04 s $\pm$ 0.49 s, respectively.

For the rest of the discussion in the paper, we refer to this unified two-stream 1-D CNN network as the body action interface for simplicity.

## 4 EXPERIMENT

### 4.1 Introduction

The goal of the experiment is to evaluate the proposed body action interface in real-time and compare its performance and user preference to a conventional interface based on the Oculus touch controller. Through this experiment, we also wish to demonstrate the potential and the utility of the body action interface for interactive VR experience, particularly as a VR workout application.

### 4.2 Design

We adopted a within-subject design for the experiment. There were two blocks for an experimental session, which consisted of a body action block and a controller block. In each block, there were one practical trial for participants to get familiar with the interface and three experimental trials for data analysis. The practice trial and experimental trials are identical, which require participants to locomote from the start to the end to complete practice or actual experiments.

We designed a third-person VR locomotion task in the setting of a small town using Unreal 4.26, shown in Figure 5. The design of the experience was inspired partly by the user study by Vijayakar and Hollerbach [28], which asked participants to avoid side walls during linear locomotion and partly by the Nintendo Ring Fit Adventure, which requires players to avoid various challenging obstacles while moving. The resource of the virtual town "Assetsville Town" was obtained from the Unreal Marketplace. The avatar model and the animation clips were obtained from Adobe Maximo. The virtual camera was placed behind the back of the avatar at a distance of 3 m in depth, facing the avatar's back. The animation of the avatar was implemented using a state machine and was played when the corresponding action was recognized by the body action interface or triggered by the buttons and the thumbstick on the controller interface. The three types of obstacles, shown in Figure 6, that a participant needed to avoid included ten low wood gates (inner dimension: 1.8 m (W) × 1.05 m (H) × 0.2 m (D)), ten low wood blocks (inner dimension: 2.8 m (W) × 0.25 m (H) × 0.01 m (D)) and ten wood walls (inner dimension: 2.8 m (W) × 1.5 m (H) × 0.2 m (D)) for a single trial/run. Before each trial, these obstacles were placed in a random order along the track with a distance of 8 m. The total length of the track was 244 m (the distance from the last obstacle to the finish line was 4 m), with a width of 5.6 m. The track that a participant needed to run through had two lanes. Each had a width of 2.8 m. These obstacles required participants to squat down to slide through low wood gates, jump in place to jump over wood blocks and strafe left or right to avoid wood walls to complete the task. Participants were asked to switch to the other lane if they encountered wood walls during locomotion to avoid collision. This enabled us to test strafing movements. Participants also have the freedom to use their preferred speed to walk or jog to reach the destination. The walking speed estimation algorithm is implemented based on the algorithm by Tregillus and Folmer [28], which estimates the step period from the linear acceleration data from a VR headset. A user's intended walking speed is estimated using linear interpolation based on the estimated step period. To demonstrate the utility of head gestures, we added functionality to let participants start their task once they nodded their head when they use the body action interface for the task.

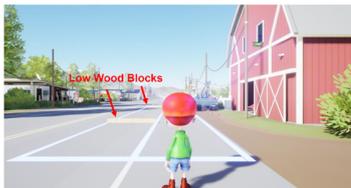

Figure 5: Screenshot of the locomotion task. An avatar stands in front of the two-lane track ready to start moving. Low wood blocks are visible in this screenshot.

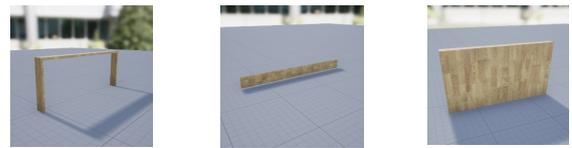

(a) Low wood gate  (b) Low wood block  (c) Wood wall

Figure 6: Three types of obstacles placed along the track.

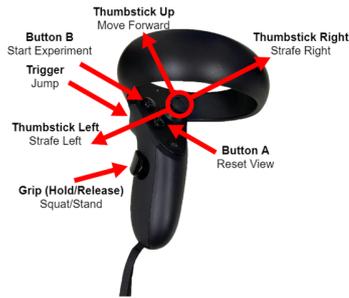

Figure 7: Thumbstick and button mapping for the Oculus touch controller. Pushing the thumbstick forward controls forward locomotion speed. Pushing the thumbstick left or right enables strafing. The trigger button triggers jumping. Holding and releasing the grip button enable squatting followed by standing. Pressing Button A resets the viewpoint and pressing Button B starts an experiment.

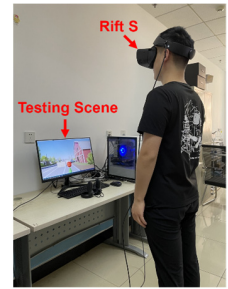

Figure 8: Experimental setup. A participant wearing an Oculus Rift S headset stands in front of the host computer, which shows the testing scene that the participant is viewing through the headset.

To compare the proposed body action interface, we designed a comparison interface based on the right Oculus touch controller, which is illustrated in Figure 7. Pushing the joystick on the right controller forward controls the forward moving speed (input scale: 0.0-1.0). A threshold of 0.5 was used to determine whether a user intended to walk ($< 0.5$) or run ($\geq 0.5$) and it also controls the switching of walking and running animations on the avatar. For the body action interface, this animation switching is controlled by recognized stepping in place action and jogging in place action. This input scale also controls the locomotion speed by multiplying the input scale with the maximum locomotion speed that is set in Unreal. Pushing the joystick left or right enables the avatar to switch to the left lane or the right lane by strafing. Holding and releasing the grip button on the right touch controller enables squatting down and standing up. Jumping is triggered by the trigger button on the right touch controller. Animation clips were played on the avatar when these actions were triggered by the touch controller.

The maximum locomotion speed for the body action interface and the controller interface was limited to 16 km/h, which corresponds to the fast-running speed of a person.

### 4.3 Participants

We invited twenty-four students from China Agricultural University (age: 18-25, 15 males and 9 females) to participate in this experiment. All had normal or corrected-to-normal vision. Participants had approximately 6-h weekly gaming time on average. The types of gaming devices used were computers and cell phones. Twenty-three participants were first-time VR users while one participant had previous VR experience. An informed consent was completed before the experiment.

### 4.4 Procedure

During an experimental session, a participant was first introduced to the task and completed the informed consent. The participant was assigned one of the two interfaces: the body action interface or the controller interface, for the first experimental block. They then wore the Oculus Rift S headset and stood in front of the host computer, shown in Figure 8. The researcher then started the VR locomotion application and informed the participant that they could start. The participant then nodded their head for the body action condition (pressing Button B for the controller condition) to initiate a 3-s countdown. Following the countdown was a prompt indicating "experiment starts" shown in the view of the HMD. Once the countdown was over, data recording was started immediately. Participants then used the assigned interface to perform in-place walking until they reached the finish line. After the participants completed four trials with the current given interface, they were directed to complete the user interface questionnaire, shown in Table 3. They then took a short rest and completed the second experimental block with another four trials using the other interface.

### 4.5 Metrics

#### 4.5.1 Subjective Metrics

The subjective metrics to evaluate the body action interface and the controller interface was based on the user interface questionnaire, shown in Table 3, adapted from previous studies [6],

Table 3: User Interface Questionnaire.

| | |
|---|---|
| 1. | The interface is easy to learn. |
| 2. | The interface is easy to use. |
| 3. | The interface is natural and intuitive to use. |
| 4. | The interface helps make the task fun. |
| 5. | Using the interface is tiring. |
| 6. | The interface helps me respond quickly. |
| 7. | The interface helps me make accurate responses. |

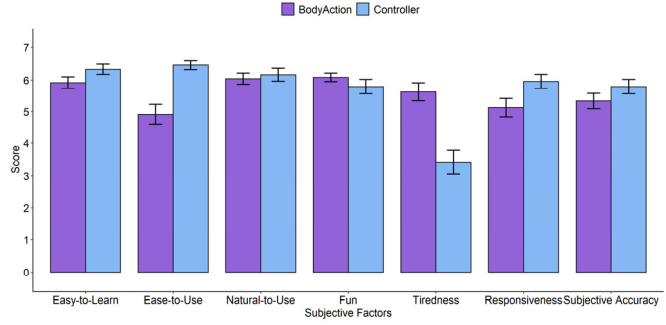

Figure 9: Results of the subjective factors of the user interface. Error bars denote the standard error of the mean.

[29], [30]. The factors that were evaluated included ease-to-learn, ease-to-use, natural-to-use, fun, tiredness, responsiveness and subjective accuracy. The questionnaire was rated using a seven-point Likert scale.

### 4.5.2 Objective Metrics

We define the following objective metrics based on [30]:
- **Task completion time $t_c$**: The duration from the start of the locomotion task to the instant when participants reached the destination to finish the task.
- **Average locomotion speed $s_l$**: The average locomotion speed computed by dividing the total length of the locomotion path by task completion time $t_c$.
- **Total number of collisions with obstacles $n_c$**: The number of collisions of the avatar with the obstacles.
- **Total number of successfully passed obstacles $n_p$**: The number of obstacles that the avatar successfully passed or avoided.

In addition, we introduced three additional factors to evaluate users' movements based on the design of this locomotion task:
- **Number of jumps $n_j$**: The number of jumps performed during locomotion.
- **Number of running slides $n_s$**: The number of running slides triggered by squatting down during locomotion.
- **Number of lane changes $n_l$**: The number of lane changes triggered by strafing right or strafing left during locomotion.

### 4.6 Results

Statistical analyses were performed using R 4.2.1. The independent factor for the analyses was the interface type: (a) BodyAction and (b) Controller. The dependent factors were the objective metrics and subjective metrics, defined in Section 4.5.

In Figure 9, we present the results of the mean and the standard error of the mean on subjective parameters. Wilcoxon rank sum tests (Package rstatix in R) on subjective parameters revealed that there were significant effects on Ease-to-Use ($p < 0.001$, effect size $r = 0.54$), Tiredness ($p < 0.001$, effect size $r = 0.56$) and Responsiveness ($p = 0.03$, effect size $r = 0.31$). Combined with the examination of Figure 9, we concluded that the controller interface was significantly easier to use compared to the body action interface. This is reasonable as the controller interface was more conventional and only required finger movements from participants to perform actions. The body action interface, on the other hand, required the whole body of participants to be involved to trigger actions for the avatar. It can also be concluded that the body action interface was significantly physically demanding to complete the locomotion task compared to the controller interface as the body action interface involved several full body actions, such as jumping and jogging in place to complete the task. In addition, the locomotion task had a relatively long track and was more challenging compared to similar previous studies. This can also explain why the body action interface was tiring. Finally, the controller interface was significantly more responsive than the body action interface. Two reasons can be attributed to this result. First, triggering avatar movements using the controller interface only required finger movements. The body action interface, on the other hand, requires a participant to be fully involved with full-body actions to trigger the avatar's movements, as we have already discussed. Second, the body action interface has an average latency of approximately 0.56 s for body actions that do not have thresholds (discussed in Section 3.4). This resulted in a slight delay between a user's movement and the triggered avatar's movement. This latency is longer compared to the latency between pressing buttons or pushing the thumbstick and the execution of the movements of an avatar. Thus, participants felt that the body action interface was not that responsive compared to the controller interface. No significant effects were found on factors that include Ease-to-Learn ($p = 0.07$, effect size $r = 0.26$), Natural-to-Use ($p = 0.46$, effect size $r = 0.11$), Fun ($p = 0.55$, effect size $r = 0.09$) and Subjective Accuracy ($p = 0.15$, effect size $r = 0.21$). In Figure 9, we observed that the average score on Ease-to-Learn was slightly higher for the controller interface, but the effect was not significant. This indicated that the body action interface was relatively easy to learn through the explanation by the researcher before the experiment started and with the practice trial to get familiar with the body action interface. The controller interface was found to be more natural to use than the body action interface, but again the effect was not significant. In terms of fun, the body action interface had a slightly higher score. We speculated that if the task was less physically demanding and was designed with more fun elements, the effect could be significant. Finally, for subjective accuracy, participants felt that the controller interface was slightly more accurate than the body action interface. This can be explained by the fact that the body action interface relied on the two-stream 1-D CNN network for body action recognition, which cannot achieve an accuracy of 100% in recognizing body actions. The controller interface was more reliable in triggering avatar movements.

In Figure 10, we present the mean and the standard error of the mean of all objective parameters. To determine the appropriate statistical test on objective parameters, we first applied the Shapiro-Wilk's test on all objective parameters to test the normality of all data. Results showed that the distributions of all objective parameters were significantly different from the normal distribution. Thus, a non-parametric test should be applied to these data. Since we only had two conditions (a) BodyAction and (b) Controller for the independent factor, we chose to apply Wilcoxon rank sum tests on all objective parameters. Results of Wilcoxon rank sum tests showed that there were significant effects on task completion time ($p < 0.001$, effect size $r = 0.84$) and average

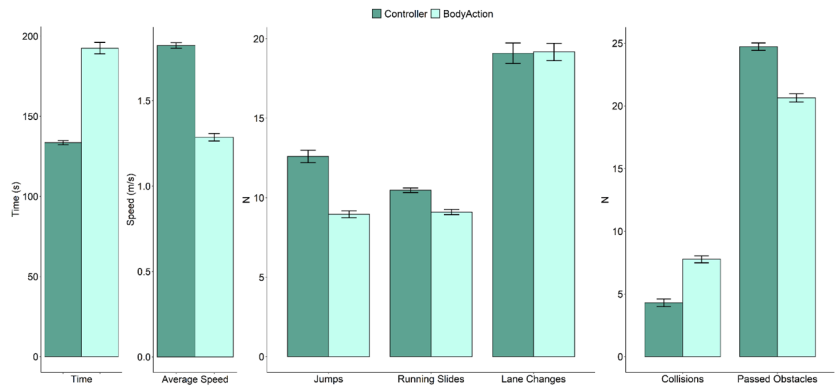
Figure 10: Results of the objective factors from the user study. Bar plot convention is as in Figure 9.

locomotion speed ($p < 0.001$, effect size $r = 0.84$). Since the body action interface is physically demanding, it can be observed from Figure 10 that completing the locomotion task required significantly more time when using the body action interface compared to using the controller interface. The average locomotion speed was also significantly lower for the body action interface than for the controller interface. Significant effects were also found on the number of jumps ($p < 0.001$, effect size $r = 0.62$), the number of running slides ($p < 0.001$, effect size $r = 0.491$) but not on the number of lane changes ($p = 0.55$, effect size $r = 0.05$). Participants had higher numbers of jumps and running slides when they used the controller interface because it was easy to trigger these actions with finger movements. However, the difference in the number of lane changes was minimal for both interfaces and the values were almost equal. This showed that lane changes were easy to perform for both interfaces. Finally, there were significant effects on the number of collisions ($p < 0.001$, effect size $r = 0.59$) and the number of successfully passed obstacles ($p < 0.001$, effect size $r = 0.64$). As shown in Figure 10, using the controller interface resulted in a lower number of collisions and a higher number of successfully passed obstacles than using the body action interface. This showed that it was significantly easier for participants to use the controller interface to pass obstacles compared to using the body action interface.

## 5 DISCUSSION

The present body action interface is a high-fidelity interface compared to the controller interface, which is low fidelity. Using a body action interface to interact with virtual environments requires participants to be fully engaged with full-body physical actions. Much energy and strength are needed to use this interface and complete the parkour-like VR locomotion task, which was long and challenging. The controller interface, on the other hand, required little effort. The control of an avatar can be done by just using finger movements. A similar finding was also reported in previous research that a controller interface outperformed a full-body tracking interaction interface based on a Kinect sensor in terms of performance [3]. Considering the efforts that one has to spend when using the body action interface, the body action interface with the parkour-like locomotion task can be intended as a VR fitness application while the controller interface can target casual-style VR locomotion gameplay.

Currently, our proposed network can only recognize head gestures while a user standing still. One challenging problem that remains to be solved is to recognize head gestures while a user performs in-place actions, such as stepping in place and jogging in place, to enable head gesture interaction while a user performs walking in place. This would require the design of novel filters or deep learning networks to separate the head motion signal component and body motion signal component from the head tracking data from an HMD. A modified 1-D CNN network or other architectures also need to be proposed to recognize in-place body actions and head gestures. Another challenging problem is to include body actions: body rotating left and body rotating right and design a method that can differentiate between body rotating left and head rotating left, and between body rotating right and head rotating right. Finally, our current design focused on in-place body actions. To include more complex body actions or body gestures related to upper limb motions, tracking data from the touch controllers should be taken into account to develop algorithms that can recognize complex body gestures.

Due to the design of the locomotion task, we only tested a subset of the body actions and head gestures that the two-stream 1-D CNN network can recognize. However, our proposed two-stream 1-D CNN network is intended as a new interface for VR interaction experience design. We expect that with the exploration and the design of novel VR experiences, other subsets or the entire set of the in-place body actions and head gestures that the network can recognize would be used in future VR application design. A limitation of the experimental design is that the current experiment only compares the proposed interface with a joystick interface. Additional experiments should consider comparing this interface to other existing approaches that can perform either body action recognition or head gesture recognition. Comparisons also need to be made with other WIP techniques to evaluate their performance and user preference.

## 6 CONCLUSION

In this paper, we presented a unified two-stream 1-D CNN network for body action and head gesture recognition. We demonstrated its utility through a VR locomotion task and compared its performance and user preference with a conventional touch controller interface. Results show that the full-body interface, as a high-fidelity interface, is physically demanding. It is less preferred compared to a low-fidelity controller interface as more effort is needed to use the body action interface. Nonetheless, this body action interface would enable the design of new VR interaction experiences with virtual environments and with avatars as it can recognize a larger number of in-place body actions and head gestures compared to existing methods and requires no specialized hardware and/or additional trackers other than an HMD. This body action interface combined with a parkour-style locomotion task is promising to be intended as VR fitness application. Other new VR interaction applications and experiences using this body action interface await further exploration.